# Interlayer Registry Determines the Sliding Potential of Layered Metal Dichalcogenides: The case of *2H*-MoS$_2$

*Adi Blumberg,[1] Uri Keshet,[1] Inbal Zaltsman,[1] and Oded Hod[*]*

Department of Chemical Physics, School of Chemistry, the Sackler Faculty of Exact

Sciences, Tel-Aviv University, Tel-Aviv 69978, Israel

We provide a simple and intuitive explanation for the interlayer sliding energy landscape of metal dichalcogenides. Based on the recently introduced registry index (*RI*) concept, we define a purely geometrical parameter which quantifies the degree of interlayer commensurability in the layered phase of molybdenum disulphide (*2H*-MoS$_2$). A direct relation between the sliding energy landscape and the corresponding interlayer registry surface of *2H*-MoS$_2$ is discovered thus marking the registry index as a computationally efficient means for studying the tribology of complex nanoscale material interfaces in the wearless friction regime.

* Corresponding author (odedhod@tau.ac.il).

---

[1] These authors contributed equally to this study.

Nanotribology is the science of friction, wear, and lubrication occurring at nanoscale interfaces. Such interfaces often appear in nanoelectromechanical systems (NEMS) which present the ultimate miniaturization of electro-mechanical devices. One of the main known caveats of NEMS is their low mechanical durability resulting from severe effects of friction and wear on systems which are characterized by high surface-to-volume ratio. While, in principle, lubrication should reduce such effects, traditional liquid phase lubricants usually fail to perform under nanoscale confined conditions as they become too viscous. Thus, one of the primary goals of nanotribology is the design of new materials that will present low friction at the atomic level.

Recent experiments on pristine solid-state layered materials have shown strong dependence of their interlayer sliding friction on the misfit angle where friction was found to nearly vanish when sliding occurred out of registry. This unique phenomenon, termed superlubricity, marks layered materials as promising candidates for serving as active components in nanoelectromechanical systems (NEMS)[1-10] as well as improved solid lubricants for macroscopic devices.[11-17]

Among the various members of the family of layered materials $MoS_2$ and $WS_2$ have been long known to serve as excellent solid lubricants.[13,18-20] Despite the wide spread use of these materials as lubrication additives the experimental[17,21-22] and theoretical[18-19,23-25] study of the nanoscopic origin of their tribological behavior remains a very active field of research. Theoretical studies often rely either on molecular dynamics simulations based on appropriately parameterized force fields[18] or on first-principles calculations using advanced density functional theory approximations.[23,25] While such calculations usually result in remarkable agreement with experimental measurements their level of complexity often shadows the relation between atomic scale processes and collective tribological material properties. In the present paper, we derive a simple and intuitive geometrical model which enables the characterization of the sliding energy landscape of *2H*-$MoS_2$ and directly relates it to the detailed atomic structure of the system.

To this end, we utilize the registry index (RI) concept[26] which has recently proven to be an efficient and reliable tool for quantifying the registry mismatch in bilayer systems and mimicking their corrugated sliding energy landscape.[26-28] Within this approach each atom in the unit cell is ascribed with a circle centered around its position and the overlaps between the projections of circles assigned to atoms located on one layer with circles associated with atoms belonging to the other layer are calculated. The obtained overlaps are appropriately summed to produce a simple numerical measure of the overall registry mismatch. This numerical value is then normalize to the range of [0:1] where 0 represents perfect interlayer registry and 1 stands for the worst stacking mode in terms of the total energy. In the case of graphene and *h*-BN where the sliding between two perfect two-dimensional layers is considered, a single circle is ascribed to each atomic position.[26-27] Here, the situation is somewhat more complex as each $MoS_2$ layer is composed of three parallel sub-layers (see Fig. 1). Therefore, the choice of circle radius has to reflect the distance between each pair of sub-layers considered. To this end, we ascribe to each atomic position two radii representing the different interactions between sub-layers 2,3 and 1',2' as described in Fig. 1.[29] We mark these radii as $r_\alpha^\beta$ where $\alpha$ is the atom around which the circle is centered and $\beta$ is the corresponding atom on the other layer (see Fig 2). As will be shown below, in order to obtain optimal fitting with density functional theory (DFT) calculations the following radii are chosen: $r_S^S = 0.9\,\text{Å}$, $r_S^{Mo} = 0.8\,\text{Å}$, and $r_{Mo}^S = 0.3\,\text{Å}$.[30]

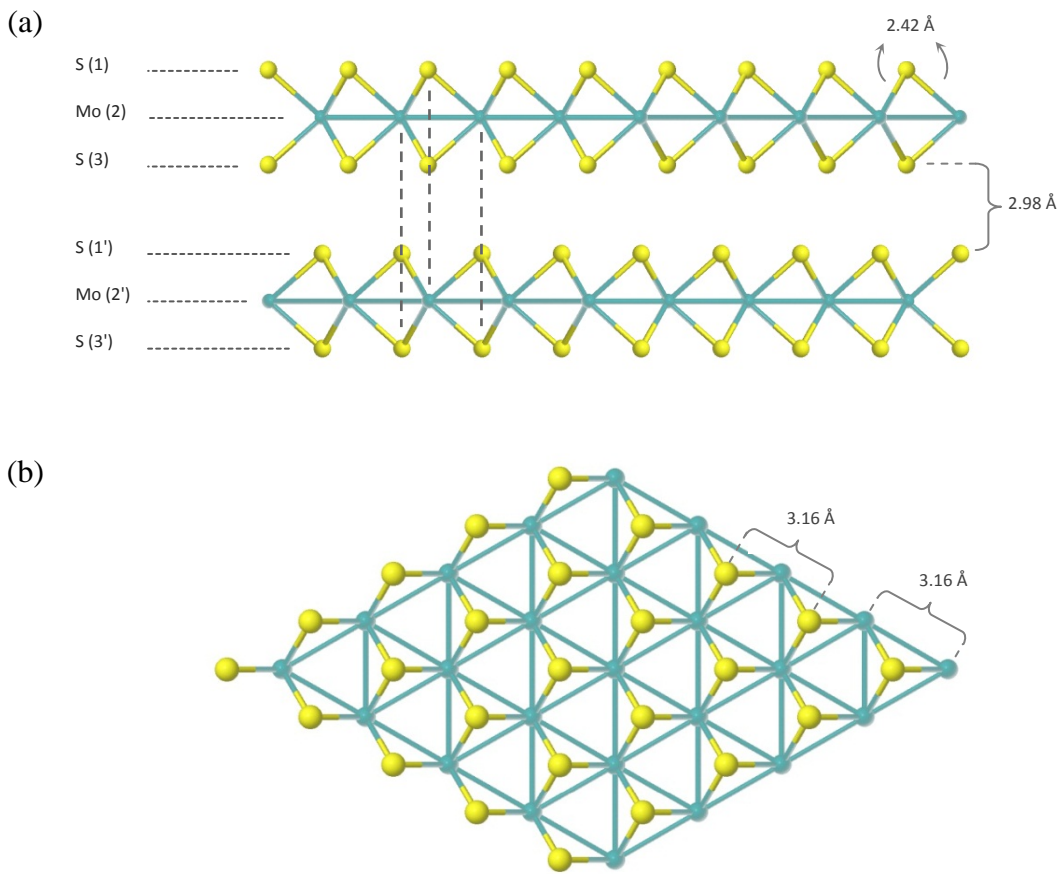

**Fig. 1** Crystal structure of MoS$_2$: (a) Side view of two layers in the optimal *AA'* stacking mode emphasizing the structure of the three sub-layers. (b) Top view of a single layer showing the hexagonal crystal symmetry.

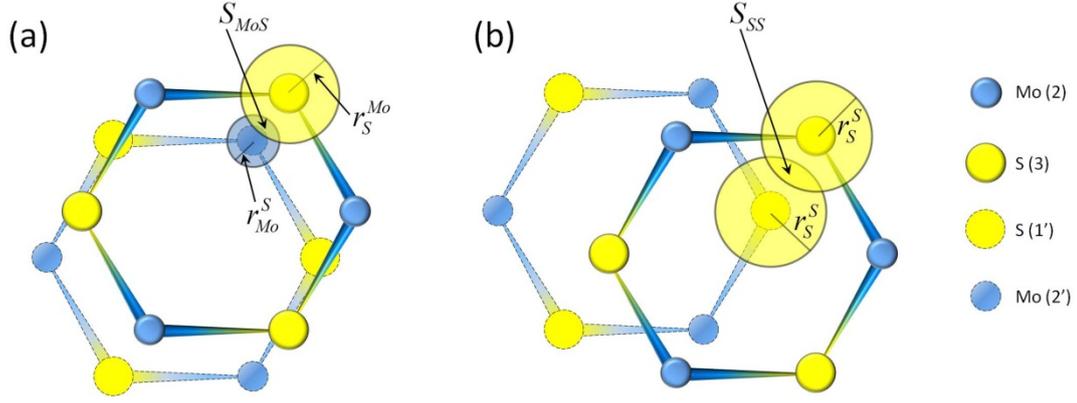

**Fig. 2** Projected circles overlaps used in the definition of the registry index: (a) $S_{MoS}$- projected overlap between a circle of radius $r_{Mo}^{S}$ centered on a Mo atom of one layer and a circle of radius $r_{S}^{Mo}$ centered on a S atom of the adjacent layer; (b) $S_{SS}$- projected overlap between two circles of radii $r_{S}^{S}$ centered on two S atoms belonging to adjacent layers.

For the definition of the registry index the optimal and worst (in terms of energy) interlayer stacking modes have to be identified. The optimal (most energetically stable) interlayer configuration is known to be the $AA'$ stacking mode where S atoms of one layer reside atop Mo atoms of the other layer (see Fig. 3(a)). [23] Starting from this configuration, the worst (highest in energy) laterally shifted interlayer configuration is the $AB_2$ stacking mode where the positions of S atoms from both layers are fully eclipsed and the Mo atoms reside above the centers of the hexagons of the adjacent layers (see Fig. 3(c)). We note that the S-S overlap ($S_{SS}$ - see Fig. 2) is maximal at the worst stacking mode whereas the Mo-S overlap ($S_{MoS}$ - see Fig. 2) is maximal at the optimal stacking mode. Therefore, Since we would like the registry index to be maximal at the worst staking mode and minimal and the optimal stacking mode, we set it to be proportional to $RI \propto (S_{SS} - S_{MoS})$. Finally, normalizing this expression to the range [0:1] yields:

$$RI = \frac{(S_{SS} - S_{SS}^{AA'}) - (S_{MoS} - S_{MoS}^{AA'})}{(S_{SS}^{AB_2} - S_{SS}^{AA'}) - (S_{MoS}^{AB_2} - S_{MoS}^{AA'})}$$

where $S_{SS}^{AA'}$, and $S_{MoS}^{AA'}$ are the S-S and Mo-S overlaps at the $AA'$ stacking mode, respectively, and $S_{SS}^{AB_2}$, and $S_{MoS}^{AB_2}$ are the S-S and Mo-S overlaps at the $AB_2$ stacking mode, respectively.

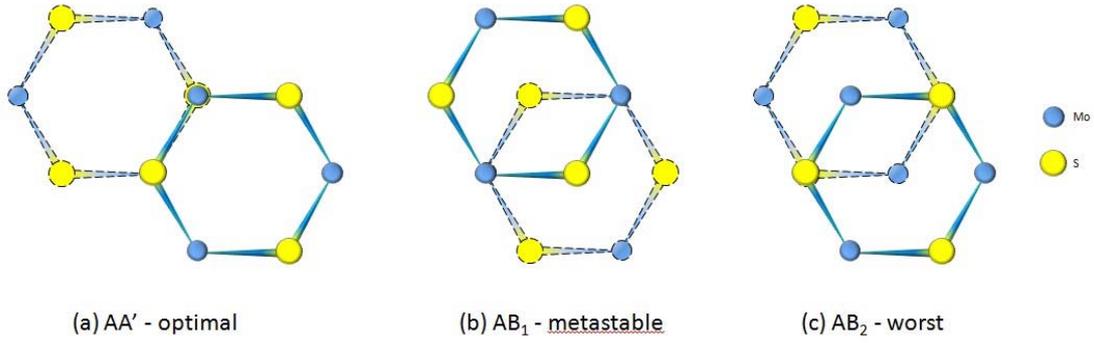

**Fig. 3** High symmetry stacking modes of MoS$_2$: (a) *AA'* configuration – the optimal stacking mode; (b) *AB$_1$* configuration – a metastable stacking mode; (c) *AB$_2$* configuration – the worst stacking mode.

Once we have a closed expression for the *RI* we can calculate it for various interlayer configurations and compare the resulting *RI* surface to the sliding energy landscape obtained from DFT calculations. To this end, we construct a unit cell of bilayer *2H*-MoS$_2$ using the lattice parameters of the bulk crystal (see supplementary material). The unit-cell of the lower layer is multiplied to form a sufficiently large finite sheet and a single unit-cell of the upper layer is then shifted with respect to the finite sheet to represent different relative interlayer positions. At each interlayer position the *RI* is recalculated resulting in a full *RI* sliding surface.

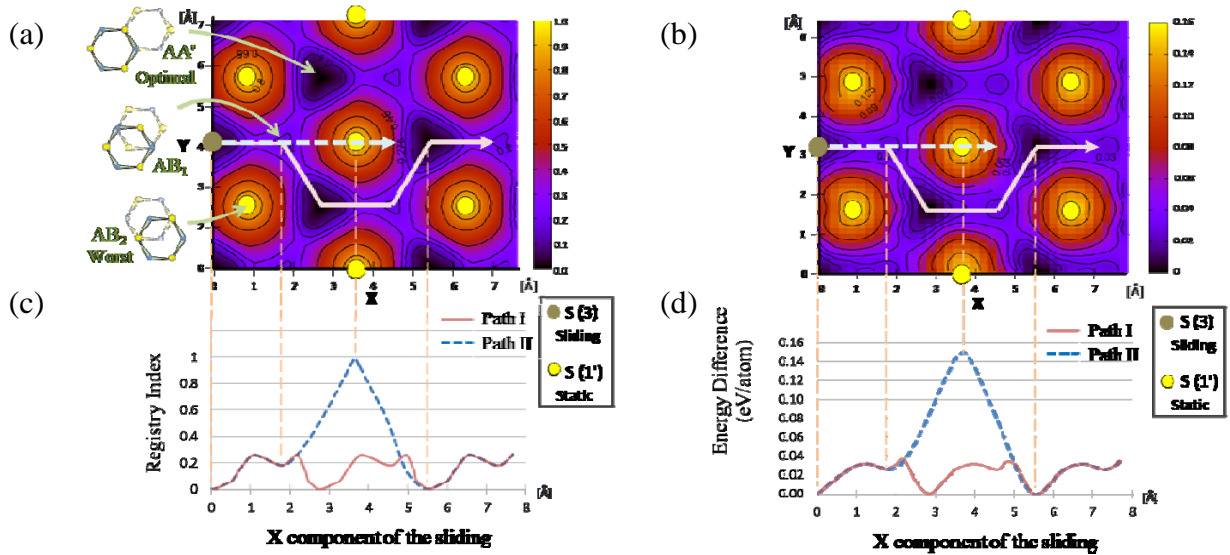

**Fig. 4:** Interlayer sliding energy landscape of bilayer *2H*-MoS2. Top: (a) registry Index and (b) total DFT energy variations (eV/atom) as a function of lateral interlayer displacements in the X-Y plane. Bottom: (c) registry index and (d) total DFT energy variations (eV/atom) along specific sliding pathways. DFT results, under an external pressure of 500 MPa, were reproduced with the kind permission and help of Prof. Simon R. Phillpot and his co-workers from Tao Liang, W. Gregory Sawyer, Scott S. Perry, Susan B. Sinnott, and Simon R. Phillpot, "First-principles determination of static potential energy surfaces for atomic friction in MoS$_2$ and MoO$_3$", *Physical Review B* **77**, 104105 (2008). Copyright (2008) by the American Physical Society.

In Fig. 4 we present the main result of this study where the sliding energy landscape obtained under an external pressure of 500 MPa using DFT calculations at the local density approximation level of theory [23] is compared with the predictions of the *RI* model. First we consider the total registry index landscape (Fig. 4(a)) as compared to the full sliding energy surface obtained via the DFT calculations (Fig. 4(b)). As can be seen, remarkable agreement between the two surfaces is achieved. The simple *RI* model is able to fully capture all important physical features appearing in the sliding energy landscape including all stationary points which occur at high symmetry interlayer configurations. By construction the *RI* model correctly predicts the *AA'* stacking mode to be the lowest energy interlayer configuration and the *AB₂* mode to be the highest energy stacking mode. Furthermore, the *AB₁* stacking mode, where the positions of Mo atoms from both layers are fully eclipsed and the S atoms reside at the centers of the hexagons of the adjacent layers (see Fig. 3(b)), is found to be a local minimum on the registry index landscape in accordance with its metastable nature obtained via the DFT results. To better appreciate the agreement between the two models we present in Fig. 4(c) slices of the full *RI* landscape along specific pathways passing through the different surface minima and maxima. The choice of pathways was directed to match that presented in ref [29]. For both paths considered excellent agreement between the *RI* variations and the DFT energy changes (Fig. 4(d)) is obtained.

The effect of external pressure on the sliding physics of *2H*-MoS$_2$ can be evaluated by comparing the results presented above to recent DFT calculations of the sliding energy landscape of this material under an external pressure of 15 GPa.[25,31] At this higher external pressure the repulsions between electron clouds of atoms belonging to two adjacent layers are considerably enhanced. As can be seen in Fig. 5(b), this is manifested in larger variations of the sliding energy landscape resulting in higher energetic barriers for interlayer sliding. Furthermore, the asymmetry observed between the *AA'* and the *AB₁* stacking modes at lower external pressure is almost completely removed in the higher pressure calculation. Interestingly, by an appropriate choice of the circle radii the registry index landscape can be tuned to reproduce the higher pressure calculations results (see Fig. 5(a)). This exemplifies the flexibility of the *RI* method for describing the sliding physics in layered materials under different external conditions. Furthermore, comparing the radii of the different atomic circles used to reproduce the two DFT calculations provides insights as for the origin of the effects of the external pressure on the overall sliding energy landscape. Specifically, to produce Fig. 5 we keep $r_S^S = 0.9\,\text{Å}$ and reduce the other radii to $r_S^{Mo} = 0.15\,\text{Å}$, and $r_{Mo}^S = 0.1\,\text{Å}$. This shows that upon increasing the external load strong Pauli repulsions between the overlapping electron clouds of neighboring sulfur atoms on adjacent layers become the dominant factor determining the sliding energy landscape whereas the interactions between the more remote S-Mo sub-layers become relatively less important. This further explains the relative reduction in asymmetry between the *AA'* and *AB₁* stacking modes where the leading interlayer terms result from S-Mo and Mo-Mo overlaps rather than the dominant S-S repulsions.

The notable agreement obtained between the *RI* results and more sophisticated first-principles calculations for *2H*-MoS$_2$ further enhances our confidence that sliding energy landscapes of layered materials, however complex they may be, can be well captured by simple models based on geometrical considerations. These, in turn, may provide an intuitive description of the physical processes underlying wearless friction in complex layered structures as well as an efficient means to study tribological properties of large scale systems which are beyond the reach of modern first-principles and molecular dynamics methods.

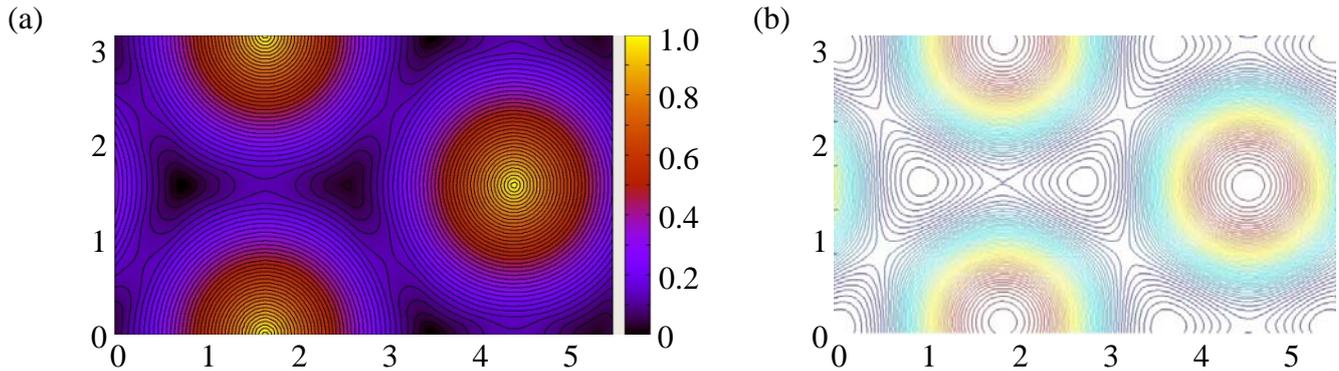

**Fig. 5:** *2H*-MoS$_2$ interlayer registry index surface (a) tuned to match the DFT sliding energy landscape at an external pressure of 15 GPa (b). The overall corrugation of the DFT sliding energy landscape is ~1 eV. DFT results were reproduced with the kind permission and help of Prof. S. Ciraci and his co-workers from S. Cahangirov, C. Ataca, M. Topsakal, H. Sahin, and S. Ciraci, "Frictional Figures of Merit for Single Layered Nanostructures", *Phys. Rev. Lett.* **108**, 126103 (2012). Copyright (2012) by the American Physical Society.

To summarize, in this work we have expanded the concept of the registry index toward complex layered materials focusing on the layered phase of MoS$_2$. Unlike the case of few layered graphene and hexagonal boron-nitride,[26-27] where each layer is composed of a single flat sheet of atoms, *2H*-MoS$_2$ has an intricate sub-layer structure which reflects on its tribological properties. Despite the involved network of interlayer interactions between different sub-layers, the registry index is able to provide an accurate description of the interlayer sliding physics in *2H*-MoS$_2$ at a fraction of the computational cost of first-principles calculations. Based on these results and the experience accumulated with the registry index thus far we believe that our suggested model can be applied for other members of the family of metal dichalcogenides, such as the layered phase of WS$_2$, and may also be expanded to describe their tubular counterparts.[26] The conclusions of the present work further affirms the robustness of the RI concept as an intuitive, flexible, and computationally efficient tool for studying nanoscale tribological characteristics of complex layered structures at the wearless friction regime.

**Acknowledgments**

The authors would like to thank Prof. Simon R. Phillpot, Tao Liang, W. Gregory Sawyer, Scott S. Perry, and Susan B. Sinnott, from the Department of Materials Science and Engineering and the Department of Mechanical and Aerospace Engineering at University of Florida, Gainesville, USA, for generously sharing the results of their DFT calculations [23] on which Figs. 4(b) and 4(d) of this paper are based. We would further like to thank Prof. S. Ciraci, S. Cahangirov, C. Ataca, M. Topsakal, and H. Sahin from UNAM-National Nanotechnology Research Center, the Institute of Materials Science and Nanotechnology, and the Department of Physics of Bilkent University, Ankara, Turkey for generously sharing the results of their DFT calculations [25] on which Fig. 5(b) of this paper is based. This work was supported by the Israel Science Foundation under grant No. 1313/08, the Center for Nanoscience and Nanotechnology at Tel Aviv University, and the Lise Meitner-Minerva Center for Computational Quantum Chemistry.

## Supplementary material

### Coordinates of some representative structures

2*H*-MoS$_2$ is a layered crystal, where each layer consists of three sub-layers of S-Mo-S (Fig. S1). The crystal unit cell has the symmetry of the P6$_3$/mmc space group, with lattice parameters of $a = b = 3.1612$ Å and $c = 12.2985$ Å.[32-33,34] For the construction of the bilayer system studied in the present manuscript we used the following partial coordinates representation:[35-36]

Mo:  (1/3)*A*+(2/3)*B*+(1/4)*C*

S:   (2/3)*A*+(1/3)*B*-(1/2-u)*C*

S:   (2/3)*A*+(1/3)*B*+(1-u)*C*

Mo:  (2/3)*A*+(1/3)*B*+(3/4)*C*

S:   (1/3)*A*+(2/3)*B*+u*C*

S:   (1/3)*A*+(2/3)*B*+(1.5-u)*C*

Where: $A = \left(-\sqrt{3}\frac{a}{2}, \frac{a}{2}, 0\right)$, $B = \left(\sqrt{3}\frac{b}{2}, \frac{b}{2}, 0\right)$, $C = (0, 0, c)$, $u = 0.621$

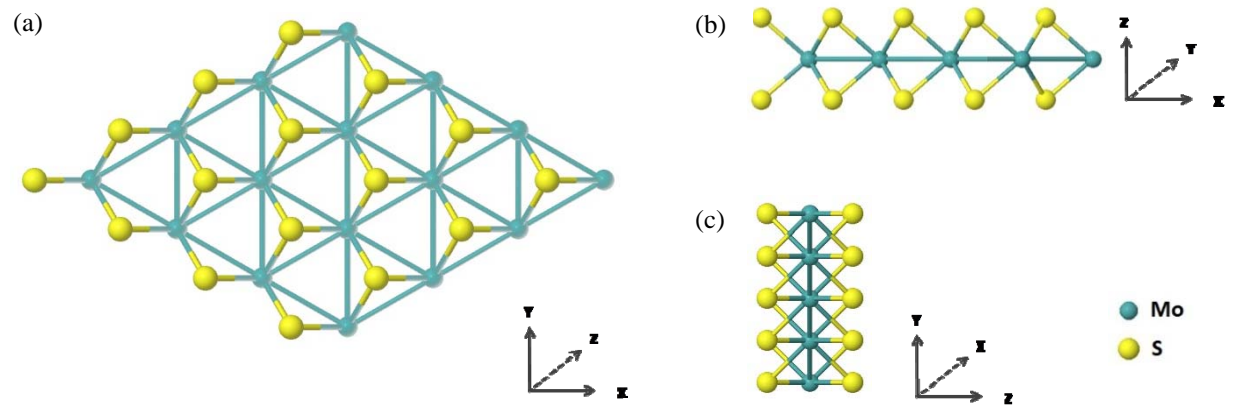

**Fig. S1** Different views of MoS$_2$ crystal structure: (a) X-Y plain, (b) Y-Z plain and (c) X-Z plain.

The resulting coordinates of some high symmetry configurations (see Fig. 3 of the main text) are given below:

## AA' configuration:

| | | | |
|---|---|---|---|
| S  | (−0.912560, | 4.741800, | 1.488119) |
| Mo | (0.912560,  | 4.741800, | 3.074625) |
| S  | (−0.912560, | 4.741800, | 4.661132) |
| S  | (0.912560,  | 4.741800, | 7.637369) |
| Mo | (−0.912560, | 4.741800, | 9.223875) |
| S  | (0.912560,  | 4.741800, | 10.810382) |

**Fig. S2** (a) AA'

## AB$_1$ configuration:

| | | | |
|---|---|---|---|
| S  | (−0.912560, | 4.741800, | 1.488119) |
| Mo | (0.912560,  | 4.741800, | 3.074625) |
| S  | (−0.912560, | 4.741800, | 4.661132) |
| S  | (2.737680,  | 4.741800, | 7.637369) |
| Mo | (0. 912560, | 4.741800, | 9.223875) |
| S  | (2.737680,  | 4.741800, | 10.810382) |

**Fig. S2** (b) AB$_1$

## AB$_2$ configuration:

| | | | |
|---|---|---|---|
| S  | (−0.912560, | 1.580600, | 1.488119) |
| Mo | (0.912560,  | 1.580600, | 3.074625) |
| S  | (−0.912560, | 1.580600, | 4.661132) |
| S  | (1.837043,  | 6.321606, | 7.637369) |
| Mo | (0.011923,  | 6.321606, | 9.223875) |
| S  | (1.837043,  | 6.321606, | 10.810382) |

**Fig. S2** (c) AB$_2$